\documentclass[prb,preprint,superscriptaddress]{revtex4}
\usepackage{amsmath,amssymb}
\usepackage{braket,graphicx}

\newcommand{\lihoyfx}{LiHo\(_x\)Y\(_{1-x}\)F\(_4\)}
\newcommand{\lihoyfag}{LiHo\(_{0.045}\)Y\(_{0.955}\)F\(_4\)}
\newcommand{\hcf}{H_\mathrm{cf}}
\newcommand{\mub}{\mu_\mathrm{B}}
\newcommand{\gl}{g_\mathrm{L}}
\newcommand{\ea}{\textit{et al.}}
\newcommand{\ho}{Ho\(^{3+}\)}

\newcommand{\ex}[1]{\langle#1\rangle}

\DeclareGraphicsExtensions{.eps}

\begin{document}

\title{Contribution of spin pairs to the magnetic response in a
  dilute dipolar ferromagnet}

\author{C.\ M.\ S.\ Gannarelli}
\affiliation{London Centre for Nanotechnology and Department of Physics and
Astronomy, University College London, Gower Street, London, WC1E 6BT,
United Kingdom}
\author{D.\ M.\ Silevitch}
\affiliation{The James Franck Institute and Department of Physics, The
University of Chicago, Chicago, Illinois 60637, USA}
\author{T.\ F.\ Rosenbaum}
\affiliation{The James Franck Institute and Department of Physics, The
University of Chicago, Chicago, Illinois 60637, USA}
\author{G.\ Aeppli}
\author{A.\ J.\ Fisher}
\email{andrew.fisher@ucl.ac.uk}
\affiliation{London Centre for Nanotechnology and Department of Physics and
Astronomy, University College London, Gower Street, London, WC1E 6BT,
United Kingdom}

\date{\today}

\begin{abstract}
We simulate the dc magnetic response of the diluted dipolar-coupled Ising magnet \lihoyfag\ in a transverse field, using exact diagonalization of a two-spin Hamiltonian averaged over nearest-neighbour configurations. The pairwise model, incorporating hyperfine interactions, accounts for the observed drop-off in the longitudinal (c-axis) susceptibility with increasing transverse field; with the inclusion of a small tilt in the transverse field, it also accounts for the behavior of the off-diagonal magnetic susceptibility.  The hyperfine interactions do not appear to lead to qualitative changes in the pair susecptibilities, although they do renormalize the crossover fields between different regimes.  Comparison with experiment indicates that antiferromagnetic correlations are more important than anticipated based on simple pair statistics and our first-principles calculations of the pair response.  This means that larger clusters will be needed for a full description of the reduction in the diagonal response at small transverse fields.
\end{abstract}

\maketitle

\section{Introduction}

The dipolar rare-earth magnetic salt LiHoF\(_4\) orders at 1.53 K\cite{bitko:1996,chakraborty:2004} to form a
Ising-like ferromagnet with long, needle-shaped domains oriented along the
Ising \(c\)-axis\cite{Cooke:1975}. This system, and the dilution series \lihoyfx\ with the magnetic \ho\ ions replaced by non-magnetic Y\(^{3+}\),   have been studied for
more than three decades\cite{Beauvillain:1978,Mennenga:1984,reich:1987,Giraud:2001,ghosh:2002,ghosh:2003,Giraud:2003,Schechter:2005,Silevitch:2007a,Quilliam:2007,jonsson:2007,Biltmo:2008,Tabei:2008,wu:1991,silevitch:2007}. 
For moderate dilution ($x>30\%$) the system continues to behave as an Ising ferromagnet \cite{Beauvillain:1978,Mennenga:1984,bitko:1996,Silevitch:2007a}; however for smaller $x$ it appears to form a spin glass at low temperatures \cite{wu:1991}.  At $x=4.5\%$ there is evidence for a novel \textit{antiglass}\cite{reich:1987} in which the scaled distribution of relaxation times \textit{loses} its low-frequency tail as the sample cools.  In this phase the material exhibits macroscopically long-lived magnetic
excitations\cite{ghosh:2002} and a novel combination of strong features in the specific heat with a featureless magnetic susceptibility which can only explained by positing long-range spin entanglement\cite{ghosh:2003}.  Some recent experiments have reported contrasting results---notably a featureless specific heat from $x=1.8\%$ to $x=8\%$  \cite{Quilliam:2007}, suggesting that the conventional spin glass may persist to lower concentrations.

The dynamics in these dilute phases are particularly interesting and could well be the key to understanding the seemingly contradictory experiments.  As well as the long-lived magnetic oscillations revealed by hole-burning experiments at $x=4.5\%$ \cite{ghosh:2002}, cotunnelling of the electronic and nuclear moments on pairs of neighboring \ho\ ions has been observed at $x=0.1\%$ \cite{Giraud:2003} through its effect on the low-frequency zero-field susceptibility.  It is appropriate to revisit the low-frequency susceptibility for several reasons.  First,  \lihoyfx\ is expected to be a model for a wide class of transverse-field dipolar systems.  Second, the observation of long decoherence times and signatures of long-range entanglement suggest the possibility of exploiting the \ho\ ions as magnetic qubits.  Finally, one would like to understand the precise role of the competition between the collective dipolar interaction, the nuclear spin bath and other decoherence pathways in determining the dynamics of the system \cite{ronnow:2005}.  Here we combine an experimental study of the magnetic response of the dilute system as we tilt the moment away from the Ising axis under large transverse fields with a  theoretical analysis in which we average over all possible pairs.  Our purpose is to establish---quantitatively---the extend to which collective (i.e. beyond-pair) effects are important for the behavior of the $x=4.5\%$ compound by doing the mos precise possible calculations of the pair susceptibility contribution at equilibrium.  The outcome is that even for this relatively high level of dilution, the collective effects are important at low transverse fields.

We presented the experimental results and a short summary of the theoretical argument in Ref.~\onlinecite{silevitch:2007}.  This paper gives full details of the model and is structured as follows. Section \ref{sec:expt}
summarizes the experimental techniques employed and captures briefly the relevant
results; Section \ref{sec:theory} describes the techniques employed in our
calculations; Section \ref{sec:comps} sets out the computational results, comparing the susceptibilities with and without hyperfine interactions and comparing them to the measured values; and Section \ref{sec:summary} presents our conclusions.

%

\section{Susceptibility measurements}
\label{sec:expt}

\begin{figure}
  \begin{center}
    \includegraphics[clip,width=3in]{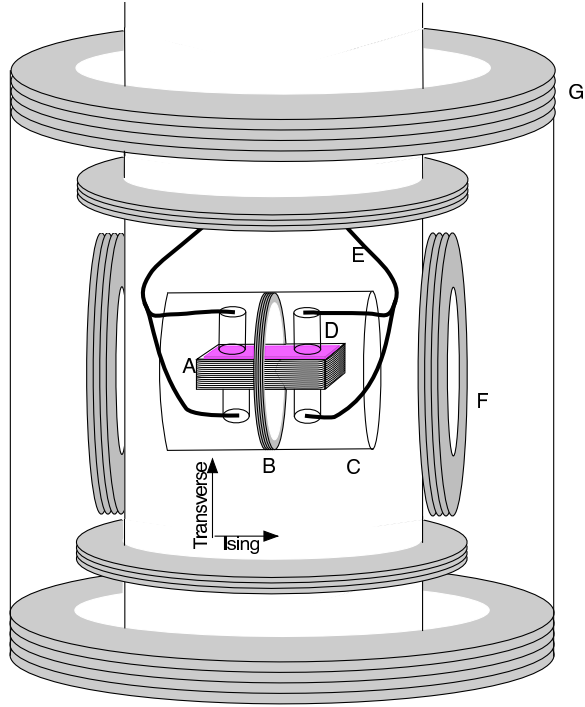}
    \caption{\label{fig:susceptometer} (Color online) Schematic of the ac vector susceptometer
      used in the experiments. The sample sits inside nested pickup coils A and B,
      sensitive to magnetic response in the transverse and Ising directions,
      respectively. An ac magnetic field along the Ising axis is supplied by
      solenoid C; the sample is thermally sunk to the cryostat cold finger via
      sapphire rods D and copper wires E. A superconducting 3-axis Helmholtz
      coil F and an 8\,T solenoid magnet G supply dc magnetic fields. G is almost, but not perfectly, aligned transverse to the $c$-axis of the sample.}
  \end{center}
\end{figure}

A single \((5\times5\times10)\)~mm\(^3\) crystal of LiHo\(_{0.045}\)Y\(_{0.955}\)F\(_4\) was characterized using ac magnetic susceptibility in a helium dilution refrigerator.
The magnetic response along the Ising axis and in the transverse plane were
experimentally measured using a specially devised multi-axis ac susceptometer, as shown in
Fig.~\ref{fig:susceptometer}. The sample was probed using a 101 Hz 2~\(\mu\)T
ac magnetic field parallel to the Ising axis. A pair of nested inductive pickup
coils allowed for simultaneous determination of the magnetic response parallel
to and transverse to the Ising axis of the crystal.   The
crystal was thermally linked to the cold finger of the refrigerator via
sapphire rods and heavy copper wires.  A multi-axis set of 100~mT Helmholtz
coils and an 8~T solenoid provided dc magnetic fields $\mathbf{H}_\mathrm{dc}$ parallel to and
almost transverse to the Ising axis respectively; however, because of the difficulty in precisely aligning the crystal, we cannot exclude the possibility of a small ($\sim 0.6^\circ$) misalignment of the solenoid from the transverse axis.  The effects of such a misalignment on the predicted properties are discussed in \S\ref{ssec:ensemble} below.

The measurement probes the diagonal and off-diagonal components respectively of the linear susceptibility tensor, but evaluated at the non-zero reference field $\mathbf{H}_\mathrm{dc}$:
\begin{eqnarray}
	\chi_{zz}&=&\left.\frac{\partial M_z}{\partial H_z}\right|_{\mathbf{H}=\mathbf{H}_\mathrm{dc}};\\
	\chi_{xz}&=&\left.\frac{\partial M_x}{\partial H_z}\right|_{\mathbf{H}=\mathbf{H}_\mathrm{dc}}.\\
\end{eqnarray}

\begin{figure}
  \begin{center}
  \includegraphics[width=3.5in]{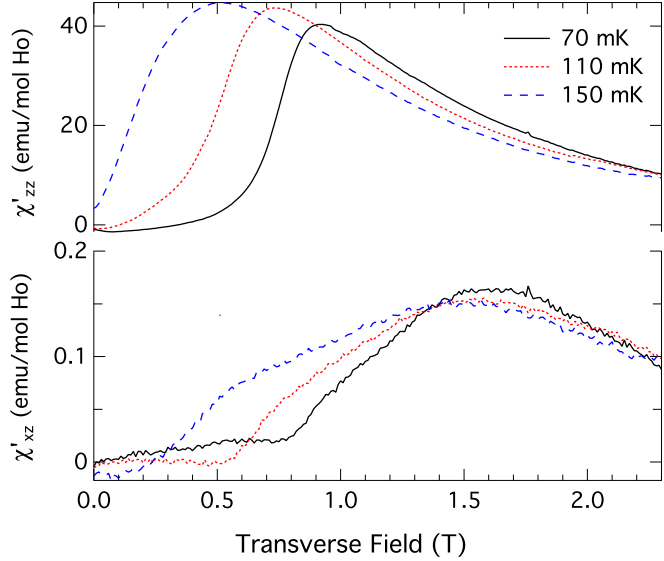} 
    \caption{\label{fig:chi_expt_1}(Color online) Measured longitudinal (top) and transverse (bottom) real
    susceptibility at 70, 110, and 150~mK (blue).  (Adapted from Ref.~\cite{silevitch:2007})}
  \end{center}
\end{figure}

Fig~\ref{fig:chi_expt_1} shows our results for the real part of the longitudinal
and transverse susceptibilities $\chi_{zz}$ and $\chi_{xz}$ as functions of $\mathbf{H}_\mathrm{dc}$ \cite{silevitch:2007}.  
These experimental results will be compared in section \ref{ssec:ensemble} to the predictions derived from the spin-pair model developed in the following sections.  The off-diagonal linear susceptibility vanishes in the limit where $\mathbf{H}_\textbf{dc}$ is exactly perpendicular to the Ising axis; as we shall see, a small component along $z$ enables $\chi_{xz}$ to capture some of the non-linear dependence of $\mathbf{M}$ on $\mathbf{H}$ and hence to give information about clustering and correlation effects, as expected from previous work \cite{Silevitch:2007a}.

The imaginary part of the magnetic response was also measured. Since the frequencies involved are small compared with all the energy scales of the microscopic Hamiltonian, a
theoretical treatment of the dissipation depends on an understanding of the
low-frequency relaxation dynamics of the Ho\(^{3+}\) ions and is not
considered in the present paper.

\section{\ho\ pair model}
\label{sec:theory}
To construct a model for the susceptibility of \ho\ pairs, we start with the complete microscopic Hamiltonian. The low-lying states of this Hamiltonian are then used to construct an effective 2-state $H$, which can be readily diagonalized for two interacting ions. If the hyperfine interactions from the microscopic single-ion Hamiltonian are added to this 2-state picture, the resulting $H$ has 16 states, and the pair Hamiltonians are still numerically tractable. Finally, a weighting scheme is implemented that incorporates contributions for pairs beyond immediate nearest-neighbors. 

\subsection{Microscopic Hamiltonian}

\begin{figure}
  \begin{center}
    \includegraphics[width=14cm]{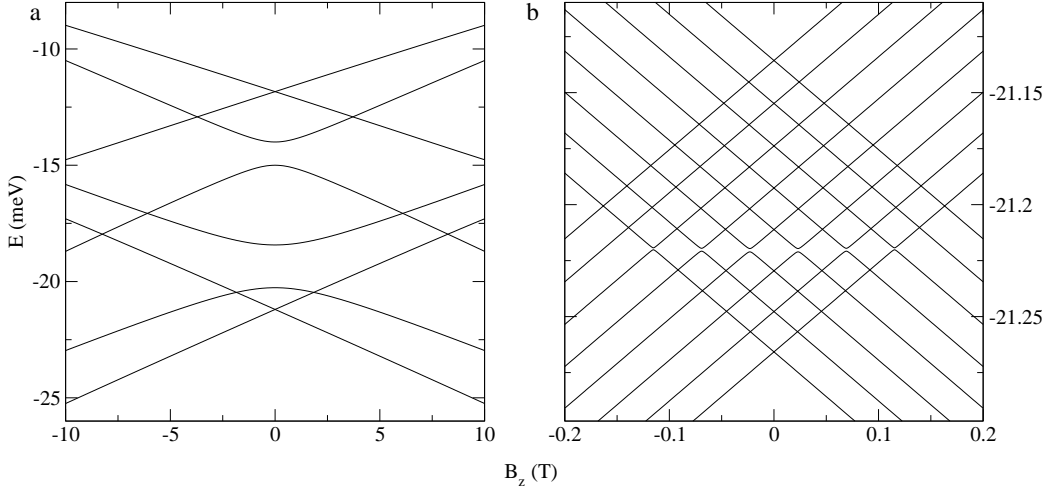}
    \caption{Single-ion energy levels as a function of longitudinal magnetic field. (a) Lowest eight electronic crystal-field levels of the
      \(^5I_8\) ground term as a function of field \(B_z\) parallel to the
      Ising axis. (b) Splitting of the two lowest electronic levels by the
      hyperfine interaction.}
    \label{fig:lev_z}
  \end{center}
\end{figure}

\begin{figure}
  \begin{center}
    \includegraphics[width=14cm]{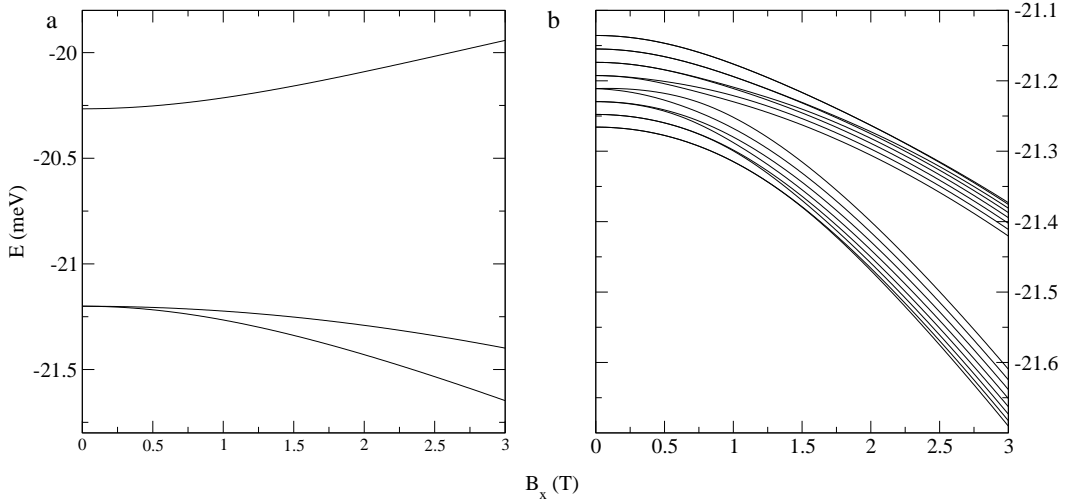}
    \caption{Single-ion energy levels as a function of transverse magnetic field. (a) Lowest three electronic crystal-field levels in the presence
      of a field \(B_x\) transverse to the Ising axis. (b) Splitting of the two lowest electronic levels by the hyperfine interaction.}
    \label{fig:lev_x}
  \end{center}
\end{figure}

The electronic Hamiltonian of a single \ho\ ion in a magnetic field is
\begin{equation}
  \label{eq:h1}
  \begin{split}
    H_1 
    &= \hcf - \mathbf{m}\cdot\mathbf{B}\\
    &=  \hcf - \mub \gl \mathbf{j}\cdot\mathbf{B}\;,
  \end{split}
\end{equation}
where \(\gl=\frac{5}{4}\) is the Land\'e \(g\) factor.
\(\hcf\) is the crystal field Hamiltonian, which splits the 17-fold
degenerate \(^5I_8\) ground term state of Ho, and is given by
\begin{equation}
  \label{eq:hcf}
  \hcf=\sum_{l=2,4,6} B_l^0 O_l^0
  +\sum_{l=4,6} B_l^4(c) O_l^4(c) + B_l^4(s) O_l^4(s)\;,
\end{equation}
where \(O_l^m\) are Stevens' operators\cite{jensen:book}. 
We follow Ref.~\onlinecite{ronnow:2007} in taking the following values for the crystal-field parameters:
\(B_2^0=-0.06\)~meV,
\(B_4^0=3.5\times 10^{-4}\)~meV,
\(B_4^4=3.6\times 10^{-3}\)~meV,
\(B_6^0=4\times 10^{-7}\)~meV,
\(B_6^4(c)=7.0\times 10^{-5}\)~meV and
\(B_6^4(s)=9.8\times 10^{-6}\)~meV. The resulting electronic energy levels are shown in
 Figs. \ref{fig:lev_z}a and \ref{fig:lev_x}a as a function
of fields parallel and transverse to the Ising axis.

The isotropic hyperfine coupling to the local \(I=\frac{7}{2}\) Ho\(^{3+}\) nuclear spin
can be included explicitly by defining
\begin{equation}
  \label{eq:hhf}
  H_\mathrm{hf} 
  = \hcf\otimes \mathbb{I}_\mathrm{N} + A \mathbf{J}\cdot\mathbf{I}
  + \mub \gl \mathbf{J}\cdot\mathbf{B} + \mu_\mathrm{N} \mathbf{I}\cdot\mathbf{B},
\end{equation}
with \(J_\alpha=j_\alpha\otimes\mathbb{I}_\mathrm{N}\) and $A/k_B=0.039\,\mathrm{K}$ or $A=3.4\,\mu\mathrm{eV}$. Figs. \ref{fig:lev_z}b and \ref{fig:lev_x}b show the effect of the hyperfine
splitting on the lowest two crystal-field states (but computed using the entire single-ion Hamiltonian (\ref{eq:hhf})).  As emphasized by Ronnow \textit{et.\ al.} \cite{ronnow:2005} and Schechter and Stamp \cite{Schechter:2005,schechter:2008}, although $A$ is small compared with the characteristic intra-ion electronic energy scales, it is comparable to the inter-ion dipolar coupling (see \S\ref{ssec:hyperfine}).  Its effect is to suppress the mixings between the two terms of the lowest electronic doublet at low temperatures, because the lowest electro-nuclear spin state in each branch has the nuclear and electronic moments anti-aligned and the nuclear moments cannot be reversed at low orders by any of the terms in equation~(\ref{eq:hhf}).

The state-space required to correctly describe the \(^5I_8\) ground
term of \ho\ in the presence of hyperfine splitting is then
\((2\times 8+1) \times (2\times \frac{7}{2} + 1) = 136\). The full Hilbert space on an ion pair therefore has dimensionality $136^2=18496$, which is inconveniently large for the repeated exact diagonalizations required to treat a range of pair geometries and fields.  We therefore proceed by truncating the model to a smaller state space while preserving the essential behavior.

\subsection{The electronic two-state system}

Following Chakraborty \ea{}\cite{chakraborty:2004}, we note the large (9.5~K)
gap between the ground state doublet and the first excited crystal-field level (Fig.\ref{fig:lev_x}a). We therefore construct
a Hamiltonian describing the low-energy behavior of the ion on a
two-dimensional electronic Hilbert space, covering only these states. This is a
parameterized model in which the inter-level repulsion shown in
Fig.~\ref{fig:lev_x}a is included explicitly as described below.

For a given value of transverse field \(B_x\) the following two-state
Hamiltonian is defined:
\begin{equation}\label{eq:twostateH}
  H^{(2)}\equiv E_0(B_x)\mathbb{I}_2 + \frac{1}{2}\Delta(B_x) \sigma_x
  +\mub\gl\mathbf{j}_\mathrm{eff}.\mathbf{B}'\;.
\end{equation}
Here \(\mathbb{I}_2\) is the identity operator in two dimensions and
\(\sigma_x\) is a spin-half Pauli operator. \(E_0(B_x)\) is the
mid-point of the lowest two energy levels and \(\Delta(B_x)\) their splitting
in that transverse field. The effective angular momentum operators are chosen
to reproduce the correct physical angular momentum matrix elements for the two
states; their decomposition into Pauli operators is discussed in
Ref.~\onlinecite{chakraborty:2004}. Finally, the field \(\mathbf{B}\) has been
replaced with \(\mathbf{B}'\equiv\mathbf{B}-B_x \mathbf{\hat{i}}\).

Note that at first sight one might expect that it would also be possible to construct a \textit{three-state} model, including the two-fold degenerate ground state as well as the first excited state, which are relatively well separated from the rest of the spectrum (see Figure~\ref{fig:lev_z}). However it turns out that level repulsion from the rest of the spectrum becomes significant at modest external fields \cite{chakraborty:2004}, and for this reason it is preferable to parameterize a two-state effective Hamiltonian operator for every value of
transverse field in order to incorporate all these effects.

In the presence of
the \(I=\frac{7}{2}\) hyperfine interaction, the two-state model becomes
\begin{equation}
  H^{(2)}_\mathrm{hf}\equiv E_0(B_x)\mathbb{I}_{16} + \frac{1}{2}\Delta(B_x) \sigma_x\otimes\mathbb{I}_8
  +\mub\gl\mathbf{J}_\mathrm{eff}\cdot\mathbf{B}'
  +\mu_\mathrm{N}\mathbf{I}\cdot\mathbf{B}+A \mathbf{J}_\mathrm{eff}\cdot\mathbf{I}\;,
\end{equation}
with \(\mathbf{J}_\mathrm{eff}\equiv\mathbf{j}_\mathrm{eff}\otimes \mathbb{I}_\mathrm{N}\).
This has a dimensionality of 16, and thus the Hamiltonian of a pair of spins
will have a numerically tractable dimensionality of 256.  In this paper we therefore retain the full nuclear Hilbert space when considering the hyperfine interaction, rather than restricting the model further to the lowest electro-nuclear doublet as in Ref.~\onlinecite{schechter:2008}.

\subsection{Intra-ion coupling}
We neglect the small exchange interactions between the \ho\ ions, so in our model pairs are coupled only by the magnetic dipole interaction.  Angular momentum operators are
constructed for each spin in a direct product Hilbert space.  The dipole
coupling between spins at \(\mathbf{R}_1\) and \(\mathbf{R}_2\) is then
\begin{equation}
  H_{12}= \frac{\mu_0(\mub\gl)^2}{R_{12}^3}\sum_{\alpha\beta}
  \left(\delta_{\alpha\beta}-\frac{3 R_{12}^\alpha R_{12}^\beta}{R_{12}^2}\right)
  J^{(1)}_\alpha J^{(2)}_\beta\;,
\end{equation}
where \(\mathbf{R}_{12}\equiv \mathbf{R}_{2}- \mathbf{R}_{1}\) and $J^{(i)}_\alpha$ is component $\alpha$ of the total angular momentum of ion $i$.  The total Hamiltonian of the pair is
\begin{equation}\label{eq:Hpair}
	H_\mathrm{pair}=H_1+H_2+H_{12}.
\end{equation}

Note that for a given pair, $H_{12}$ gives rise to an effective field at site 1
\begin{equation}
	B^{(1)}_{\mathrm{eff},\alpha}=\frac{\mu_0(\mub\gl)}{R_{12}^3}\sum_{\beta}
  \left(\delta_{\alpha\beta}-\frac{3 R_{12}^\alpha R_{12}^\beta}{R_{12}^2}\right)
   J^{(2)}_\beta,
\end{equation}
which in general contains a transverse component.  Strictly, therefore, the field-dependent parameters in equation~(\ref{eq:twostateH}) should be computed incorporating this component.  However, in practice this dependence is negligible for the applied fields of interest because the characteristic scale of $B^{(1)}$ is at most $\mu_0\mub\gl|J^{(2)}|/a^3=29\,\mathrm{mT}$, while the experimental variation of $\chi$ is on the scale of fields that can mix the Ising doublet, of order 1\,T (see Figures~\ref{fig:chi_expt_1} and \ref{fig:lev_x}).

\subsection{Computing the susceptibility}

The isothermal susceptibility is defined as 
\begin{equation}\label{eq:suscdef}
  \chi_{\alpha\beta}\equiv\frac{1}{V}\left(\frac{\partial \left<m_\alpha\right>}
    {\partial H_\beta}\right)_T\;,
\end{equation}
where $\mathbf{m}$ is the total magnetic moment and $V$ is the sample volume.
We apply this by computing the field-dependent eigenstates of the pair Hamiltonian (\ref{eq:Hpair}) and computing
\begin{equation}
  \begin{split}
    \chi_{\alpha\beta}&=-\frac{1}{k_BTZV}\sum_i\exp(-E_i/k_BT)
    \braket{i|\Delta \hat{m}_\alpha|i}\braket{i|\Delta \hat{m}_\beta|i}\\
    &\quad+\frac{1}{ZV}\sum_i\exp(-E_i/k_BT)\sum_j^{}\,^{'}
    2\Re\left[\frac{\braket{i|\hat{m}_\alpha|j}\braket{j|\hat{m}_\beta|i}}
      {E_i-E_j}\right]\\
      &=\chi_\text{Pauli}+\chi_\text{Van Vleck}\;,\\
  \end{split}
\end{equation}
where the primed sum goes over all states \(i\) and \(j\) such that \(E_i\neq
E_j\) and $\Delta m_\alpha\equiv m_\alpha-\ex{m_\alpha}$. 
Matrix elements between degenerate states have been made to vanish by
a choice of basis such that \(\hat{m}_\beta\) is diagonal in each degenerate
subspace. Numerically we assume states \(i\) and \(j\) are degenerate if
\(E_j-E_i<\varepsilon\), a small value chosen such that the susceptibility is not
sensitive to variations in \(\varepsilon\). Note that in applying equation~(\ref{eq:suscdef}) we assume that the \ho\ ions remain in thermal equilibrium over the timescales of the experiment, i.e. that all thermalizing relaxation processes operate on a timescale fast compared to the measurement.


\subsection{A pair-ensemble weighting scheme}
\label{ssec:weight}
We wish not only to
examine the behavior of specific pairs of spins, but also to calculate the average response for a distribution of spin pairs corresponding to the physical \lihoyfag\ crystal. We proceed by assuming that at this dilute concentration the behavior of each spin is affected only by the closest spin and compute a weighted susceptibility. This is determined by computing
the susceptibility of an exhaustive sample of pairs of spins up to some cutoff
distance \(r_\mathrm{c}\) and weighting each term by the probability
that in a randomly populated set of sites in a lattice with mean fractional
occupancy \(x\), the chosen spin \(s_2\) would be the nearest occupied site to
the reference spin \(s_1\). If all the sites were at different distances, this would be given by the probability that no sites nearer to \(s_1\) than \(s_2\) are
occupied, while the site \(s_2\) itself is occupied. The weighting for a site
\(s_j\) would then be
\begin{equation}
w_j=x(1-x)^{N_j}\;,
\end{equation}
where \(N_j\) is the number of sites closer to \(s_1\) than \(s_j\).
However in practice the sites $s_2$ occur in `shells' with equal distance from $s_1$; if there are $n_j$ sites in shell $j$, we ascribe a weighting to each site which is a fraction $1/n_j$ of the probability that there is at least one neighboring spin anywhere in the shell:
\begin{equation}
w_j=\left[\frac{1-(1-x)^{n_j}}{n_j}\right](1-x)^{N_j}\;.
\end{equation}

The cutoff distance \(r_\mathrm{c}\) is always chosen such that the probability of the nearest occupied site
\(s_2\) being more than \(r_\mathrm{c}\) from \(s_1\) does not significantly exceed \(10^{-3}\); the required $r_\mathrm{c}$ therefore increases as $x$ falls.  For the calculations presented here we included 22 shells of neighbors containing 146 ions,  corresponding to $r_\mathrm{c}=2.58\,a=13.4$~\AA.  At the experimental spin concentration  ($x=0.045$) the probability that the pair separation exceeds $r_\mathrm{c}$ is then $1.20\times 10^{-3}$.

\section{Results}
\label{sec:comps}

\subsection{Contributions of individual pairs}

\begin{figure}
\includegraphics[width=3.5in]{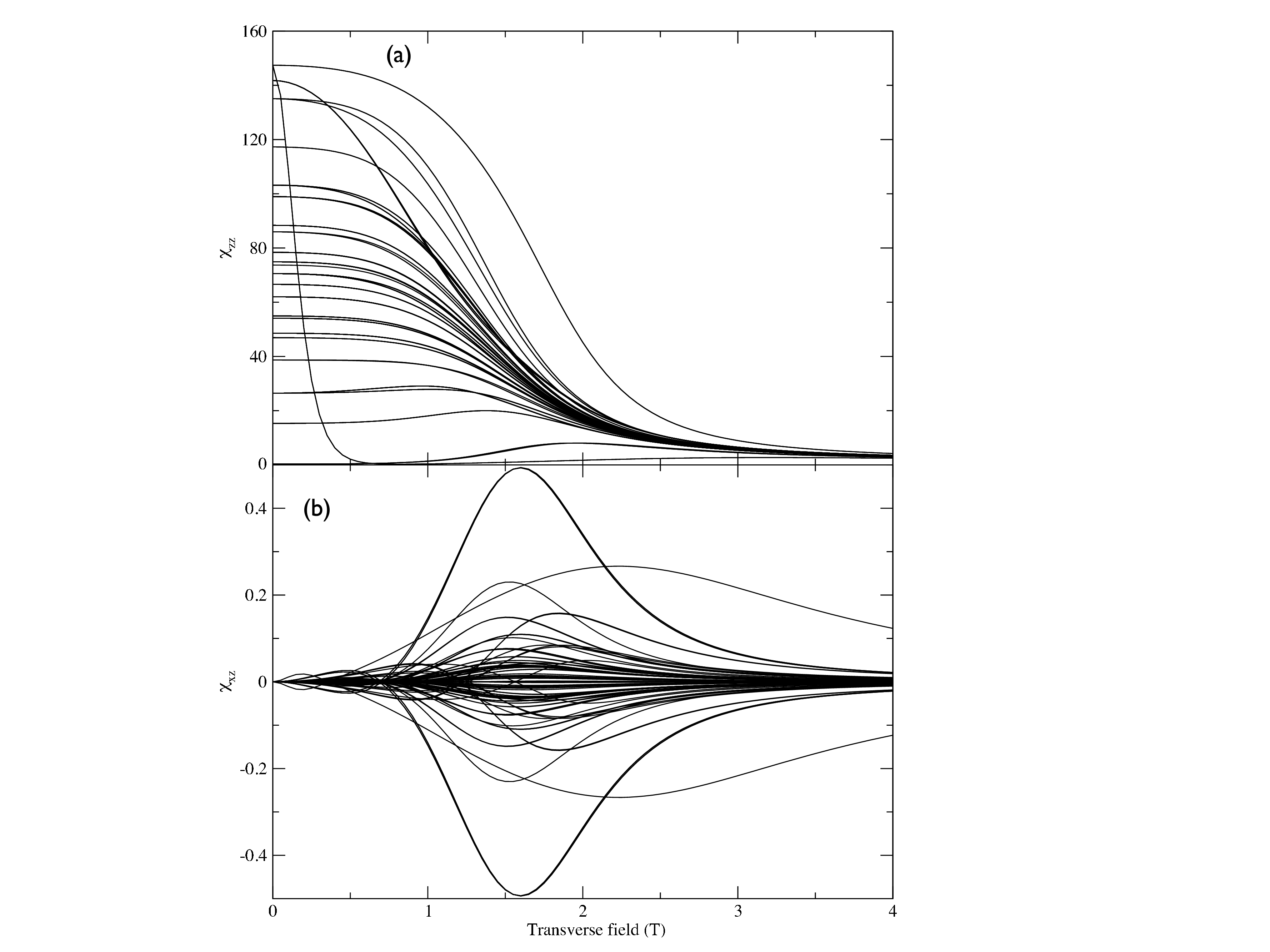}
\caption{\label{fig:chi_hf_all} Computed susceptibilities for all pairs of spins at T=70 mK, with hyperfine interactions included. (a) Diagonal response $\chi_{zz}$. (b) Off-diagonal response $\chi_{xz}$.}
\end{figure}

\begin{figure}
  \begin{center}
    \includegraphics[width=3.5in]{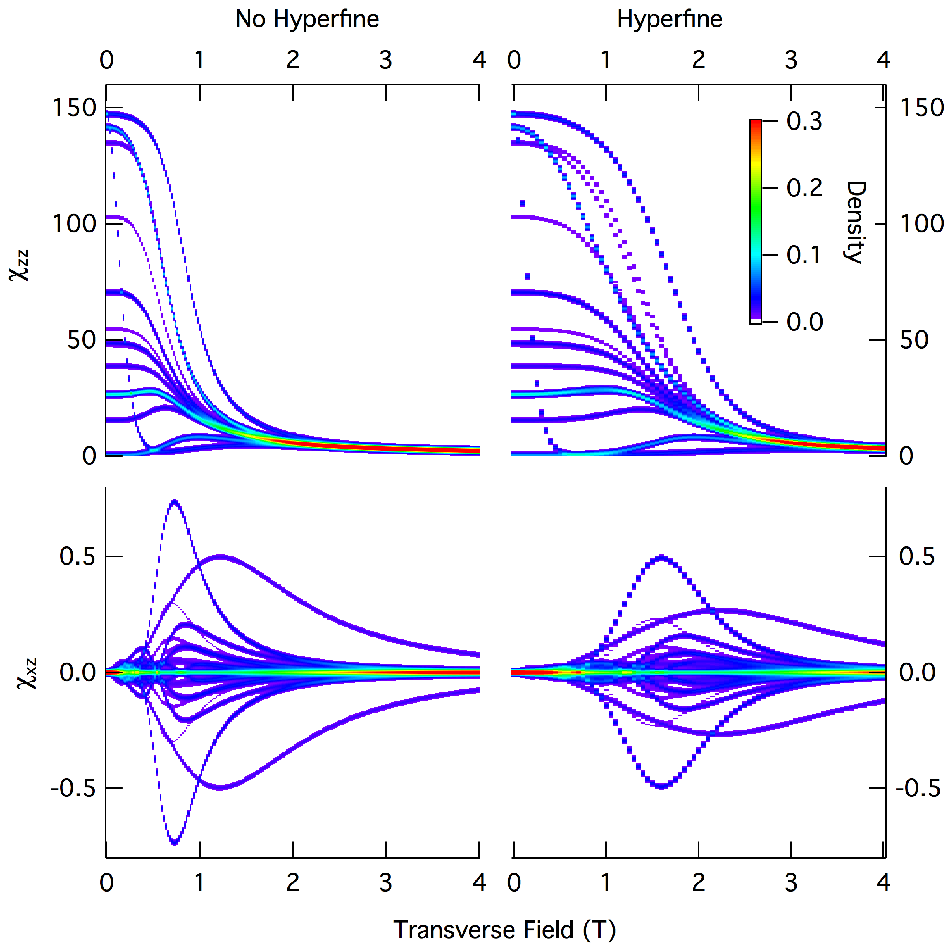}
    \caption{\label{fig:chixz_hf_dens} (Color online)
      Contribution of the various pairs to the
      ensemble-averaged functions \(\chi_{xz}(B_{x})\) (top) and
      \(\chi_{zz}(B_{x})\) (bottom). Left and right, respectively, show the effects of omitting and including the hyperfine term in the Hamiltonian.  The temperature was $T=70\,\mathrm{mk}$ and the field was applied along $(1,0,0)$.}
  \end{center}
\end{figure}

The magnetic response of a pair of Ho spins depends strongly on their
separation and orientation. Fig~\ref{fig:chi_hf_all} shows the Ising-axis and transverse response of all pairs that make a significant
contribution to the cluster ensemble. Although these plots are of illustrative value in demonstrating the wide range of
behaviors arising from spin pairs, it is more useful to examine how these
different responses contribute to the ensemble average. Fig~\ref{fig:chixz_hf_dens} shows these averages
by plotting the susceptibilities of each pair using the weighting $w_i$ as a color map. Susceptibility bands appear in this weighted map due to particular closely neighboring spin pairs.
 It can also be seen that for every pair of spins with
a transverse susceptibility \(\chi_{xz}(B_x)=f(B_x)\), there exists a pair with
\(\chi_{xz}(B_x)=-f(B_x)\). It thus follows that an ensemble average
as defined in Sec.~\ref{ssec:weight} will give a zero value of \(\chi_{xz}\)
for all values of field \(B_x\). As discussed below, the measured response is well described by a small (\(0.6^\circ\))
tilt of \(B_x\), producing a polarizing field along the Ising axis. A comparison of the weighted susceptibilities with and without the incorporation of hyperfine effects suggests that the primary effect of the hyperfine term is to renormalize the transverse field; this behavior is discussed in more detail in Section \ref{ssec:hyperfine} below.

\subsection{Pair orientation and response}

Depending on relative orientation, the dipole coupling can be either ferromagnetic or antiferromagnetic. A
ferromagnetic pair has a susceptibility \(\chi_{zz}\) which diverges in the
limit of low temperatures and zero transverse field, whereas an
antiferromagnetically coupled pair has vanishing susceptibility in the same
limit. As can be seen from Fig.~\ref{fig:chi_expt_1}, antiferromagnetic behavior dictates the measured response of
the sample of \lihoyfag, and as shown in Fig \ref{fig:chi_hf_all} certain pairs show a
qualitatively similar magnetic response.  As we shall see below, however, their contribution to the ensemble average used in this paper is not sufficient to make the overall average susceptibility agree with the measured one.

\begin{figure}
  \begin{center}
    \includegraphics[width=5.5in]{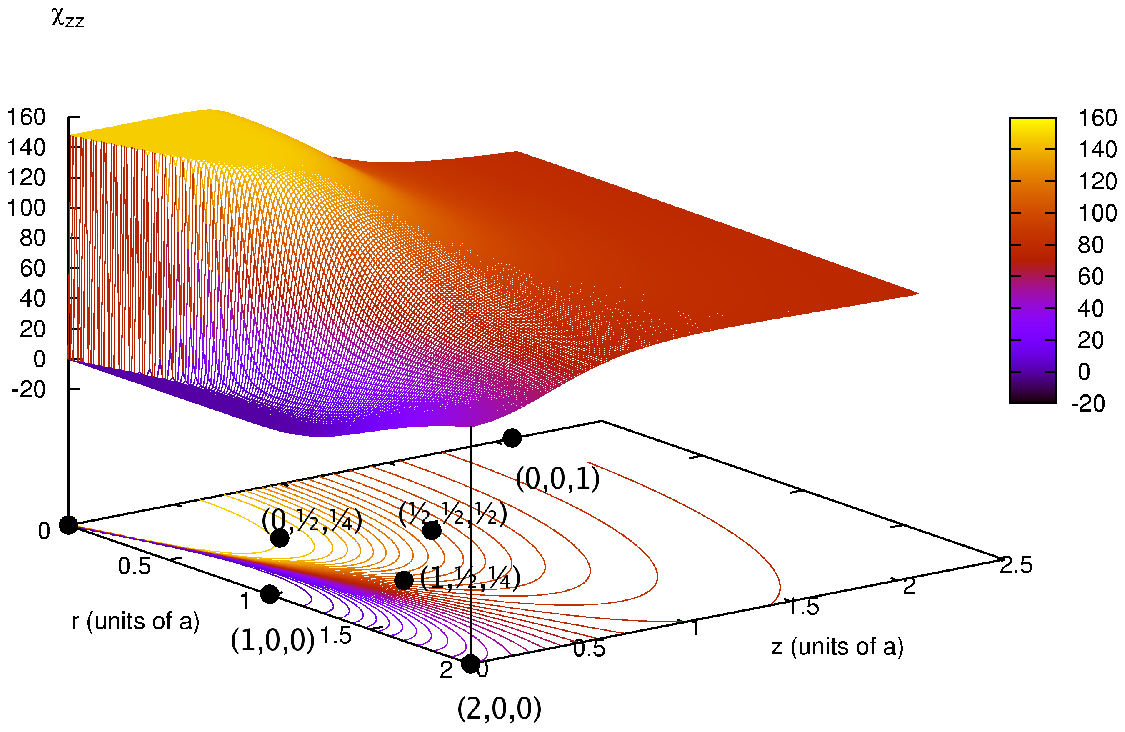}
    \caption{\label{fig:angle} (Color online)
    The effect of geometry on \(\chi_{zz}\), computed with zero transverse field and temperature $T=70\,\mathrm{mK}$. Response is plotted for a pair
    of spins with axial separation \(z\) and in-plane seperation \(r\) (units
    of lattice parameter \(a\)) with the marked points showing the locations of various nearest neighbors.  The susceptibility is shown in units of emu/mol Ho.}
  \end{center}
\end{figure}

The relation of this behavior to the crystal geometry can be understood from Fig.~\ref{fig:angle}, showing the zero-field susceptibility at $T=70\,\mathrm{mK}$ of a pair of \ho\ ions separated by a
distance \(r\) in the \(a\)--\(b\) plane and \(z\) on the \(c\)-axis.  The crossover between the ferromagnetic and antiferromagnetic couplings occurs along the line $z/r=1/\sqrt{2}$; the strongly antiferromagnetic pairs are located in-plane at
\((1,0,0)\) and \((2,0,0)\) and the most strongly ferromagnetic pair is the
nearest-neighbor pair at \( (\frac{1}{2},0,\frac{1}{4})\). Note that the
on-axis pair \((0,0,1)\) is more weakly ferromagnetic at this temperature, owing to
the larger spatial separation.

\subsection{The effect of the hyperfine interaction}
\label{ssec:hyperfine}

\begin{figure}
  \begin{center}
    \includegraphics[width=4in]{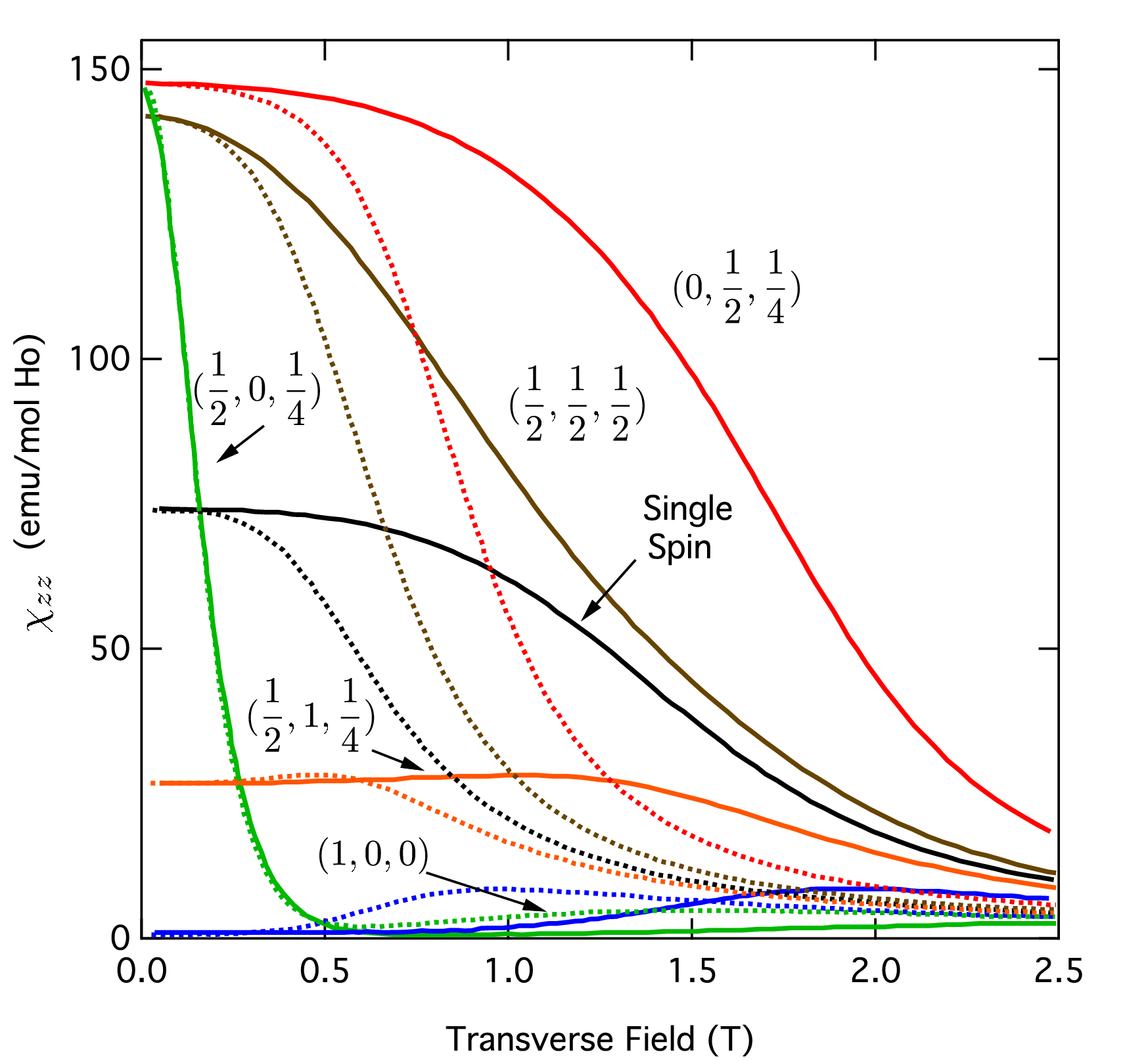}
    \caption{\label{fig:hyper} (Color online)
      The magnetic response at $T=70\,\mathrm{mK}$ of certain important spin pairs, using a Hamiltonian which incorporates hyperfine effects (solid) and which omits these effects (dotted). The primary effect of adding the hyperfine splitting is to impose an effective renormalization of the transverse field scale.  The transverse field is applied along the $(1,0,0)$ direction.}
  \end{center}
\end{figure}

We now examine the role that hyperfine interactions play in determining the behavior of the system. It is important to understand whether these effects produce a qualitative change in the behavior, as expansion of this model to $n=3$ and larger clusters of spins becomes numerically impractical if the hyperfine splittings are essential. Fig.~\ref{fig:hyper} shows susceptibilities for high-weight spin pairs both with and without hyperfine effects. (Note that pairs such as $(\frac{1}{2},0,\frac{1}{4})$ and $(0,\frac{1}{2},\frac{1}{4})$, which are equivalent at zero field, become inequivalent for non-zero fields, except when the field lies along symmetry directions such as $(1,1,0)$.) We see that the primary role of the hyperfine interactions is to renormalize the applied transverse field, rather than to introduce fundamentally different behavior. This in turn suggests that useful insights may be derived from considering larger spin clusters in the absence of the hyperfine splittings. It should be noted, however, that the strongly ferromagnetic $(\frac12,0,\frac14)$ pair does not show this renormalization when it is oriented so that the projection of the separation vector into the $ab$-plane lies along the transverse field direction.

\subsection{The ensemble-averaged susceptibilities}
\label{ssec:ensemble}

\begin{figure}
  \begin{center}
    \includegraphics[width=5.5in]{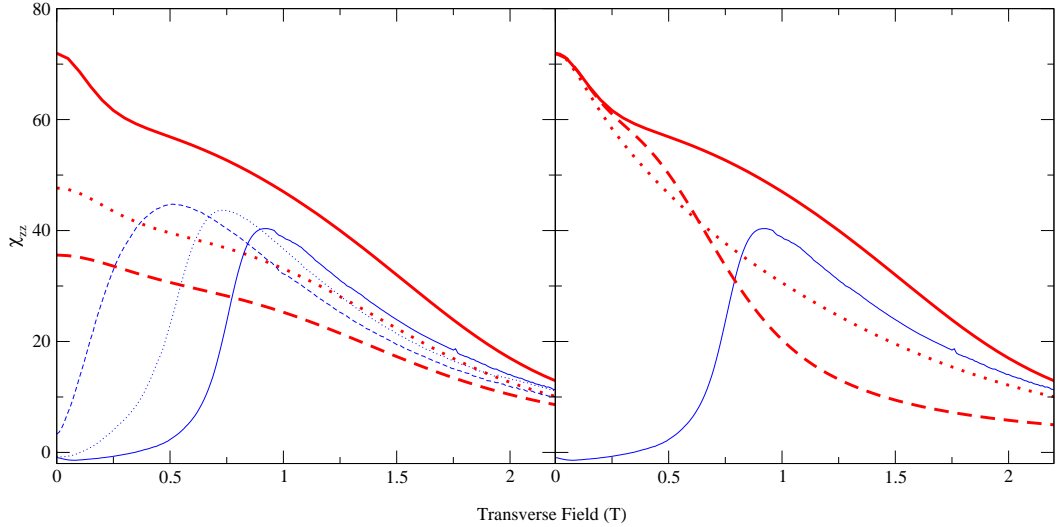}
    \caption{\label{fig:chizz_avge} (Color online)
     Measured and computed \(\chi_{zz}\) (in units of emu/mol Ho). (left) Computed (heavy, red curves) and measured (blue, light)
      susceptibility. Solid, dotted and dashed curves are T=70~mK,
      110~mK and 150~mK respectively.
      (Right) The effect of tilting the
      transverse field (dotted curve), and of omitting the hyperfine
      interaction (dashed), at T=70~mK. Unbroken red (heavy) and blue (light) curves, again
      show the computed (with hyperfine, no tilting) and measured
      susceptibilities respectively.}
  \end{center}
\end{figure}

Fig.~\ref{fig:chizz_avge} shows the experimental and ensemble-averaged
longitudinal susceptibility \(\chi_{zz}\). The left panel shows computed and
experimental results at temperatures of 70, 110 and 150~mK. Computed
results include the effect of the hyperfine response, but omit in this panel
the effect of tilting the field \(B_x\). The model captures the overall temperature dependence of the data, but it cannot account for the low-field suppression of the susceptibility because the average is dominated by the contributions of ferromagnetic and effectively uncoupled
pairs.

The right panel of Fig.~\ref{fig:chizz_avge} shows the effects of varying the parameters of the model at a constant T=70 mK. The dashed curves show the result of removing the hyperfine terms; for most of the field range, the renormalization seen in the individual pair susceptibilities is visible. At low field, the strongly ferromagnetic $(\frac12,0,\frac14)$ pairing dominates, and no renormalization is seen. The dotted curve shows the result of keeping the hyperfine effects and adding a $0.6^\circ$ tilt to the applied field, with the attendant slight polarization along the Ising axis. We can see that this improves the match between the high-field behavior of the model and the experiment. 

\begin{figure}
  \begin{center}
    \includegraphics[width=5.5in]{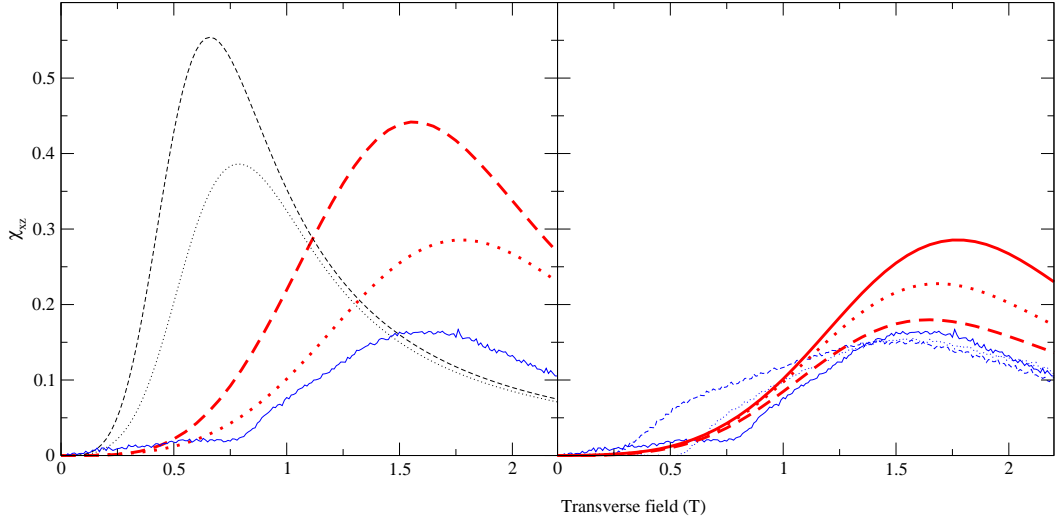}
    \caption{\label{fig:chixz_avge} (Color online)
      The apparent transverse susceptibility resulting from a transverse field
      tilted by \(0.6^\circ\). (Left) Measured
      susceptibility (blue, thin, solid curve) is contrasted with the
      susceptibility of a single ion in a tilted field (dashed curves), and 
      the pairwise average susceptibility (dotted). The heavy, red curves
      include hyperfine effects. The thin, red curves do not.
      (Right) The effect of temperature. Measurements
      are shown as light, blue curves; calculations as heavy, read curves.
      Temperatures are 70~mK (solid curve), 110~mK (dotted) and 150~mK
      (dashed).}
  \end{center}
\end{figure}

Fig.~\ref{fig:chixz_avge} displays similar information for \(\chi_{xz}\). Note that owing to the symmetry observed in Fig.~\ref{fig:chi_hf_all}(b), the
ensemble average of \(\chi_{xz}\) vanishes in the absence of a polarizing field.
Thus, the only appropriate comparison is between the tilted-field computation
and the measured value, as shown in the left pane of Fig.~\ref{fig:chixz_avge} for both
single-ion and ensemble-pair-average computations. It is clear that
the tilt is responsible for the measured effect, with the pair average providing a better match to the measured susceptibility than a single-ion calculation. The
effect of the hyperfine response is the same renormalization of the field seen in the longitudinal response.
The right pane of this figure shows the effect of temperature on both the
measured and the pairwise average \(\chi_{xz}\).

\section{Conclusions}
\label{sec:summary}

We have developed a spin-pair model for understanding the behavior of dilute \lihoyfx. A weighted ensemble average of all spin pairs reproduces the high-transverse-field experimental susceptibility, but not the low-field antiferromagnetic character of the data.  Nonetheless, the rise in the longitudinal susceptibility at a transverse field of around 1\,T, which looks like a signature of a spin gap, does correspond to the calculated susceptibility for certain antiferromagnetic pairs.  This suggests that a full understanding of the system requires treatment of larger clusters, an extension which should be numerically feasible because of the observation that the primary effect of the hyperfine splitting in the dc susceptibility is to renormalize the transverse field. This will allow extension of the model to larger clusters of spins using the simplified 2-state description for individual spins rather than a full 16-state description. Ultimately, to reach the thermodynamic limit, it would still be necessary to generalize a scaling approach, such as the real-space renormalization group of Ref.~\cite{ghosh:2003}, to include finite transverse fields.

Such an extension would sample somewhat different regions of configuration space, since Fig.~\ref{fig:angle} shows that
the antiferromagnetic region extends considerably farther in distance than does the
ferromagnetic region. This space is not sampled significantly in the pairwise model,
owing to the rapid fall-off of the weighting function \(w_i\) with distance,
but larger clusters can sample this interaction region far more extensively.


\begin{acknowledgments}The work at the University of Chicago was supported by U.S. DOE Basic Energy Sciences Grant No. DE-FG02-99ER45789, and that at UCL by the U.K. Engineering and Physical Sciences Research Council under grant EP/D049717/1.\end{acknowledgments}


\bibliography{just_pairs}

\begin{thebibliography}{22}
\expandafter\ifx\csname natexlab\endcsname\relax\def\natexlab#1{#1}\fi
\expandafter\ifx\csname bibnamefont\endcsname\relax
  \def\bibnamefont#1{#1}\fi
\expandafter\ifx\csname bibfnamefont\endcsname\relax
  \def\bibfnamefont#1{#1}\fi
\expandafter\ifx\csname citenamefont\endcsname\relax
  \def\citenamefont#1{#1}\fi
\expandafter\ifx\csname url\endcsname\relax
  \def\url#1{\texttt{#1}}\fi
\expandafter\ifx\csname urlprefix\endcsname\relax\def\urlprefix{URL }\fi
\providecommand{\bibinfo}[2]{#2}
\providecommand{\eprint}[2][]{\url{#2}}

\bibitem[{\citenamefont{Bitko et~al.}(1996)\citenamefont{Bitko, Rosenbaum, and
  Aeppli}}]{bitko:1996}
\bibinfo{author}{\bibfnamefont{D.}~\bibnamefont{Bitko}},
  \bibinfo{author}{\bibfnamefont{T.~F.} \bibnamefont{Rosenbaum}},
  \bibnamefont{and} \bibinfo{author}{\bibfnamefont{G.}~\bibnamefont{Aeppli}},
  \bibinfo{journal}{Phys. Rev. Lett.} \textbf{\bibinfo{volume}{77}},
  \bibinfo{pages}{940} (\bibinfo{year}{1996}).

\bibitem[{\citenamefont{Chakraborty et~al.}(2004)\citenamefont{Chakraborty,
  Henelius, Kj\o{}nsberg, Sandvik, and Girvin}}]{chakraborty:2004}
\bibinfo{author}{\bibfnamefont{P.~B.} \bibnamefont{Chakraborty}},
  \bibinfo{author}{\bibfnamefont{P.}~\bibnamefont{Henelius}},
  \bibinfo{author}{\bibfnamefont{H.}~\bibnamefont{Kj\o{}nsberg}},
  \bibinfo{author}{\bibfnamefont{A.~W.} \bibnamefont{Sandvik}},
  \bibnamefont{and} \bibinfo{author}{\bibfnamefont{S.~M.}
  \bibnamefont{Girvin}}, \bibinfo{journal}{Phys. Rev. B}
  \textbf{\bibinfo{volume}{70}}, \bibinfo{pages}{144411}
  (\bibinfo{year}{2004}).

\bibitem[{\citenamefont{Cooke et~al.}(1975)\citenamefont{Cooke, Jones, Silva,
  and Wells}}]{Cooke:1975}
\bibinfo{author}{\bibfnamefont{A.~H.} \bibnamefont{Cooke}},
  \bibinfo{author}{\bibfnamefont{D.~A.} \bibnamefont{Jones}},
  \bibinfo{author}{\bibfnamefont{J.~F.~A.} \bibnamefont{Silva}},
  \bibnamefont{and} \bibinfo{author}{\bibfnamefont{M.~R.} \bibnamefont{Wells}},
  \bibinfo{journal}{Journal of Physics C: Solid State Physics}
  \textbf{\bibinfo{volume}{8}}, \bibinfo{pages}{4083} (\bibinfo{year}{1975}).

\bibitem[{\citenamefont{Beauvillain et~al.}(1978)\citenamefont{Beauvillain,
  Renard, Laursen, and Walker}}]{Beauvillain:1978}
\bibinfo{author}{\bibfnamefont{P.}~\bibnamefont{Beauvillain}},
  \bibinfo{author}{\bibfnamefont{J.}~\bibnamefont{Renard}},
  \bibinfo{author}{\bibfnamefont{I.}~\bibnamefont{Laursen}}, \bibnamefont{and}
  \bibinfo{author}{\bibfnamefont{P.}~\bibnamefont{Walker}},
  \bibinfo{journal}{Phys Rev B} \textbf{\bibinfo{volume}{18}},
  \bibinfo{pages}{3360} (\bibinfo{year}{1978}).

\bibitem[{\citenamefont{Mennenga et~al.}(1984)\citenamefont{Mennenga, de~Jongh,
  and Huiskamp}}]{Mennenga:1984}
\bibinfo{author}{\bibfnamefont{G.}~\bibnamefont{Mennenga}},
  \bibinfo{author}{\bibfnamefont{L.}~\bibnamefont{de~Jongh}}, \bibnamefont{and}
  \bibinfo{author}{\bibfnamefont{W.}~\bibnamefont{Huiskamp}},
  \bibinfo{journal}{Journal of Magnetism and Magnetic Materials}
  \textbf{\bibinfo{volume}{44}}, \bibinfo{pages}{59} (\bibinfo{year}{1984}).

\bibitem[{\citenamefont{Reich et~al.}(1987)\citenamefont{Reich, Rosenbaum, and
  Aeppli}}]{reich:1987}
\bibinfo{author}{\bibfnamefont{D.~H.} \bibnamefont{Reich}},
  \bibinfo{author}{\bibfnamefont{T.~F.} \bibnamefont{Rosenbaum}},
  \bibnamefont{and} \bibinfo{author}{\bibfnamefont{G.}~\bibnamefont{Aeppli}},
  \bibinfo{journal}{Phys. Rev. Lett.} \textbf{\bibinfo{volume}{59}},
  \bibinfo{pages}{1969} (\bibinfo{year}{1987}).

\bibitem[{\citenamefont{Giraud et~al.}(2001)\citenamefont{Giraud, Wernsdorfer,
  Tkachuk, Mailly, and Barbara}}]{Giraud:2001}
\bibinfo{author}{\bibfnamefont{R.}~\bibnamefont{Giraud}},
  \bibinfo{author}{\bibfnamefont{W.}~\bibnamefont{Wernsdorfer}},
  \bibinfo{author}{\bibfnamefont{A.}~\bibnamefont{Tkachuk}},
  \bibinfo{author}{\bibfnamefont{D.}~\bibnamefont{Mailly}}, \bibnamefont{and}
  \bibinfo{author}{\bibfnamefont{B.}~\bibnamefont{Barbara}},
  \bibinfo{journal}{Phys Rev Lett} \textbf{\bibinfo{volume}{87}},
  \bibinfo{pages}{057203} (\bibinfo{year}{2001}).

\bibitem[{\citenamefont{Ghosh et~al.}(2002)\citenamefont{Ghosh, Parthasarathy,
  Rosenbaum, and Aeppli}}]{ghosh:2002}
\bibinfo{author}{\bibfnamefont{S.}~\bibnamefont{Ghosh}},
  \bibinfo{author}{\bibfnamefont{R.}~\bibnamefont{Parthasarathy}},
  \bibinfo{author}{\bibfnamefont{T.~F.} \bibnamefont{Rosenbaum}},
  \bibnamefont{and} \bibinfo{author}{\bibfnamefont{G.}~\bibnamefont{Aeppli}},
  \bibinfo{journal}{Science} \textbf{\bibinfo{volume}{296}},
  \bibinfo{pages}{2195} (\bibinfo{year}{2002}).

\bibitem[{\citenamefont{Ghosh et~al.}(2003)\citenamefont{Ghosh, Rosenbaum,
  Aeppli, and Coppersmith}}]{ghosh:2003}
\bibinfo{author}{\bibfnamefont{S.}~\bibnamefont{Ghosh}},
  \bibinfo{author}{\bibfnamefont{T.~F.} \bibnamefont{Rosenbaum}},
  \bibinfo{author}{\bibfnamefont{G.}~\bibnamefont{Aeppli}}, \bibnamefont{and}
  \bibinfo{author}{\bibfnamefont{S.~N.} \bibnamefont{Coppersmith}},
  \bibinfo{journal}{Nature} \textbf{\bibinfo{volume}{425}}, \bibinfo{pages}{48}
  (\bibinfo{year}{2003}).

\bibitem[{\citenamefont{Giraud et~al.}(2003)\citenamefont{Giraud, Tkachuk, and
  Barbara}}]{Giraud:2003}
\bibinfo{author}{\bibfnamefont{R.}~\bibnamefont{Giraud}},
  \bibinfo{author}{\bibfnamefont{A.~M.} \bibnamefont{Tkachuk}},
  \bibnamefont{and} \bibinfo{author}{\bibfnamefont{B.}~\bibnamefont{Barbara}},
  \bibinfo{journal}{Phys Rev Lett} \textbf{\bibinfo{volume}{91}},
  \bibinfo{pages}{257204} (\bibinfo{year}{2003}).

\bibitem[{\citenamefont{Schechter and Stamp}(2005)}]{Schechter:2005}
\bibinfo{author}{\bibfnamefont{M.}~\bibnamefont{Schechter}} \bibnamefont{and}
  \bibinfo{author}{\bibfnamefont{P.~C.~E.} \bibnamefont{Stamp}},
  \bibinfo{journal}{Phys Rev Lett} \textbf{\bibinfo{volume}{95}},
  \bibinfo{pages}{267208} (\bibinfo{year}{2005}).

\bibitem[{\citenamefont{Silevitch
  et~al.}(2007{\natexlab{a}})\citenamefont{Silevitch, Bitko, Brooke, Ghosh,
  Aeppli, and Rosenbaum}}]{Silevitch:2007a}
\bibinfo{author}{\bibfnamefont{D.~M.} \bibnamefont{Silevitch}},
  \bibinfo{author}{\bibfnamefont{D.}~\bibnamefont{Bitko}},
  \bibinfo{author}{\bibfnamefont{J.}~\bibnamefont{Brooke}},
  \bibinfo{author}{\bibfnamefont{S.}~\bibnamefont{Ghosh}},
  \bibinfo{author}{\bibfnamefont{G.}~\bibnamefont{Aeppli}}, \bibnamefont{and}
  \bibinfo{author}{\bibfnamefont{T.~F.} \bibnamefont{Rosenbaum}},
  \bibinfo{journal}{Nature} \textbf{\bibinfo{volume}{448}},
  \bibinfo{pages}{567} (\bibinfo{year}{2007}{\natexlab{a}}).

\bibitem[{\citenamefont{Quilliam et~al.}(2007)\citenamefont{Quilliam, Mugford,
  Gomez, Kycia, and Kycia}}]{Quilliam:2007}
\bibinfo{author}{\bibfnamefont{J.~A.} \bibnamefont{Quilliam}},
  \bibinfo{author}{\bibfnamefont{C.~G.~A.} \bibnamefont{Mugford}},
  \bibinfo{author}{\bibfnamefont{A.}~\bibnamefont{Gomez}},
  \bibinfo{author}{\bibfnamefont{S.~W.} \bibnamefont{Kycia}}, \bibnamefont{and}
  \bibinfo{author}{\bibfnamefont{J.~B.} \bibnamefont{Kycia}},
  \bibinfo{journal}{Phys Rev Lett} \textbf{\bibinfo{volume}{98}},
  \bibinfo{pages}{037203} (\bibinfo{year}{2007}).

\bibitem[{\citenamefont{J\"{o}nsson et~al.}(2007)\citenamefont{J\"{o}nsson,
  Mathieu, Wernsdorfer, Tkachuk, and Barbara}}]{jonsson:2007}
\bibinfo{author}{\bibfnamefont{P.~E.} \bibnamefont{J\"{o}nsson}},
  \bibinfo{author}{\bibfnamefont{R.}~\bibnamefont{Mathieu}},
  \bibinfo{author}{\bibfnamefont{W.}~\bibnamefont{Wernsdorfer}},
  \bibinfo{author}{\bibfnamefont{A.~M.} \bibnamefont{Tkachuk}},
  \bibnamefont{and} \bibinfo{author}{\bibfnamefont{B.}~\bibnamefont{Barbara}},
  \bibinfo{journal}{Physical Review Letters} \textbf{\bibinfo{volume}{98}},
  \bibinfo{eid}{256403} (pages~\bibinfo{numpages}{4}) (\bibinfo{year}{2007}),
  \urlprefix\url{http://link.aps.org/abstract/PRL/v98/e256403}.

\bibitem[{\citenamefont{Biltmo and Henelius}(2008)}]{Biltmo:2008}
\bibinfo{author}{\bibfnamefont{A.}~\bibnamefont{Biltmo}} \bibnamefont{and}
  \bibinfo{author}{\bibfnamefont{P.}~\bibnamefont{Henelius}},
  \bibinfo{journal}{Phys Rev B} \textbf{\bibinfo{volume}{78}},
  \bibinfo{pages}{054437} (\bibinfo{year}{2008}).

\bibitem[{\citenamefont{Tabei et~al.}(2008)\citenamefont{Tabei, Gingras, Kao,
  and Yavors'kii}}]{Tabei:2008}
\bibinfo{author}{\bibfnamefont{S.~M.~A.} \bibnamefont{Tabei}},
  \bibinfo{author}{\bibfnamefont{M.~J.~P.} \bibnamefont{Gingras}},
  \bibinfo{author}{\bibfnamefont{Y.~J.} \bibnamefont{Kao}}, \bibnamefont{and}
  \bibinfo{author}{\bibfnamefont{T.}~\bibnamefont{Yavors'kii}},
  \bibinfo{journal}{Phys Rev B} \textbf{\bibinfo{volume}{78}},
  \bibinfo{pages}{184408} (\bibinfo{year}{2008}).

\bibitem[{\citenamefont{Wu et~al.}(1991)\citenamefont{Wu, Ellman, Rosenbaum,
  Aeppli, and Reich}}]{wu:1991}
\bibinfo{author}{\bibfnamefont{W.}~\bibnamefont{Wu}},
  \bibinfo{author}{\bibfnamefont{B.}~\bibnamefont{Ellman}},
  \bibinfo{author}{\bibfnamefont{T.~F.} \bibnamefont{Rosenbaum}},
  \bibinfo{author}{\bibfnamefont{G.}~\bibnamefont{Aeppli}}, \bibnamefont{and}
  \bibinfo{author}{\bibfnamefont{D.~H.} \bibnamefont{Reich}},
  \bibinfo{journal}{Phys. Rev. Lett.} \textbf{\bibinfo{volume}{67}},
  \bibinfo{pages}{2076} (\bibinfo{year}{1991}).

\bibitem[{\citenamefont{Silevitch
  et~al.}(2007{\natexlab{b}})\citenamefont{Silevitch, Gannarelli, Fisher,
  Aeppli, and Rosenbaum}}]{silevitch:2007}
\bibinfo{author}{\bibfnamefont{D.~M.} \bibnamefont{Silevitch}},
  \bibinfo{author}{\bibfnamefont{C.~M.~S.} \bibnamefont{Gannarelli}},
  \bibinfo{author}{\bibfnamefont{A.~J.} \bibnamefont{Fisher}},
  \bibinfo{author}{\bibfnamefont{G.}~\bibnamefont{Aeppli}}, \bibnamefont{and}
  \bibinfo{author}{\bibfnamefont{T.~F.} \bibnamefont{Rosenbaum}},
  \bibinfo{journal}{Phys. Rev. Lett.} \textbf{\bibinfo{volume}{99}},
  \bibinfo{pages}{057203} (\bibinfo{year}{2007}{\natexlab{b}}).

\bibitem[{\citenamefont{Ronnow et~al.}({2005})\citenamefont{Ronnow,
  Parthasarathy, Jensen, Aeppli, Rosenbaum, and McMorrow}}]{ronnow:2005}
\bibinfo{author}{\bibfnamefont{H.}~\bibnamefont{Ronnow}},
  \bibinfo{author}{\bibfnamefont{R.}~\bibnamefont{Parthasarathy}},
  \bibinfo{author}{\bibfnamefont{J.}~\bibnamefont{Jensen}},
  \bibinfo{author}{\bibfnamefont{G.}~\bibnamefont{Aeppli}},
  \bibinfo{author}{\bibfnamefont{T.}~\bibnamefont{Rosenbaum}},
  \bibnamefont{and} \bibinfo{author}{\bibfnamefont{D.}~\bibnamefont{McMorrow}},
  \bibinfo{journal}{{Science}} \textbf{\bibinfo{volume}{{308}}},
  \bibinfo{pages}{{389}} (\bibinfo{year}{{2005}}).

\bibitem[{\citenamefont{Jensen and Mackintosh}(1991)}]{jensen:book}
\bibinfo{author}{\bibfnamefont{J.}~\bibnamefont{Jensen}} \bibnamefont{and}
  \bibinfo{author}{\bibfnamefont{A.~R.} \bibnamefont{Mackintosh}},
  \emph{\bibinfo{title}{Rare Earth Magnetism: Structures and Excitations}}
  (\bibinfo{publisher}{Clarendon Press}, \bibinfo{address}{Oxford},
  \bibinfo{year}{1991}).

\bibitem[{\citenamefont{R\o{}nnow et~al.}(2007)\citenamefont{R\o{}nnow, Jensen,
  Parthasarathy, Aeppli, Rosenbaum, McMorrow, and Kraemer}}]{ronnow:2007}
\bibinfo{author}{\bibfnamefont{H.~M.} \bibnamefont{R\o{}nnow}},
  \bibinfo{author}{\bibfnamefont{J.}~\bibnamefont{Jensen}},
  \bibinfo{author}{\bibfnamefont{R.}~\bibnamefont{Parthasarathy}},
  \bibinfo{author}{\bibfnamefont{G.}~\bibnamefont{Aeppli}},
  \bibinfo{author}{\bibfnamefont{T.~F.} \bibnamefont{Rosenbaum}},
  \bibinfo{author}{\bibfnamefont{D.~F.} \bibnamefont{McMorrow}},
  \bibnamefont{and} \bibinfo{author}{\bibfnamefont{C.}~\bibnamefont{Kraemer}},
  \bibinfo{journal}{Phys. Rev. B} \textbf{\bibinfo{volume}{75}},
  \bibinfo{pages}{054426} (\bibinfo{year}{2007}).

\bibitem[{\citenamefont{Schechter and Stamp}(2008)}]{schechter:2008}
\bibinfo{author}{\bibfnamefont{M.}~\bibnamefont{Schechter}} \bibnamefont{and}
  \bibinfo{author}{\bibfnamefont{P.~C.~E.} \bibnamefont{Stamp}},
  \bibinfo{journal}{Physical Review B (Condensed Matter and Materials Physics)}
  \textbf{\bibinfo{volume}{78}}, \bibinfo{eid}{054438}
  (pages~\bibinfo{numpages}{17}) (\bibinfo{year}{2008}).

\end{thebibliography}

\end{document}